\documentclass[journal=jacsat,manuscript=article,layout=twocolumn]{achemso}
\setkeys{acs}{keywords = true}
\usepackage[version=3]{mhchem} 
\usepackage{textcomp}  
\usepackage[hidelinks, colorlinks=true, allcolors=blue]{hyperref}
\usepackage{xcolor}
\usepackage[normalem]{ulem}
\usepackage{siunitx}
\usepackage{placeins}
\usepackage{amsmath}
\usepackage{color}
\usepackage{standalone}
\usepackage{dcolumn}
\usepackage{bm}
\usepackage{mathtools}
\usepackage{microtype}
\usepackage{etoolbox}
\usepackage{amssymb}
\usepackage{mathrsfs}
\usepackage{accents}
\usepackage[normalem]{ulem}
\usepackage{makecell}  
\let\oldmaketitle\maketitle
\let\maketitle\relax

\renewcommand{\normalsize}{\footnotesize}
\renewcommand{\Large}{\large}
\renewcommand{\large}{\normalsize}
\makeatletter
\def\@affilfont{\Large}
\makeatother

\author{Apostolos Apostolakis}
\affiliation[Institute of Physics]
{Institute of Physics,Czech Academy of Sciences, Na Slovance 2, CZ-18200, Prague, Czech Republic}  
\email{apostolakis@fzu.cz}
\author{Guillaume Aoust}
\affiliation[MIRSENSE]
{MIRSENSE,Nano-INNOV Batiment 863, 8 av de la Vauve, 91120 Palaiseau, France}
\author{Gr\'egory Maisons}
\affiliation[MIRSENSE]
{MIRSENSE,Nano-INNOV Batiment 863, 8 av de la Vauve, 91120 Palaiseau, France}
\author{Ludovic Laurent}
\affiliation[MIRSENSE \\]
{MIRSENSE,Nano-INNOV Batiment 863, 8 av de la Vauve, 91120 Palaiseau, France}
\author{Mauro Fernandes Pereira}
\affiliation[Khalifa University of Science and Technology]
{Department of Physics,Khalifa University of Science and Technology,  127788, Abu Dhabi, United Arab Emirates}
\alsoaffiliation[Institute of Physics]
{Institute of Physics,Czech Academy of Sciences, Na Slovance 2, CZ-18200, Prague, Czech Republic}
\email{mauro.pereira@ku.ac.ae}

\title[]
  {Photo-acoustic spectroscopy using a quantum cascade laser (QCL) for analysis of ammonia in water solutions}


\begin{document}

\begin{tocentry}
\centering
\includegraphics[width=8.1cm, height=3.6cm]{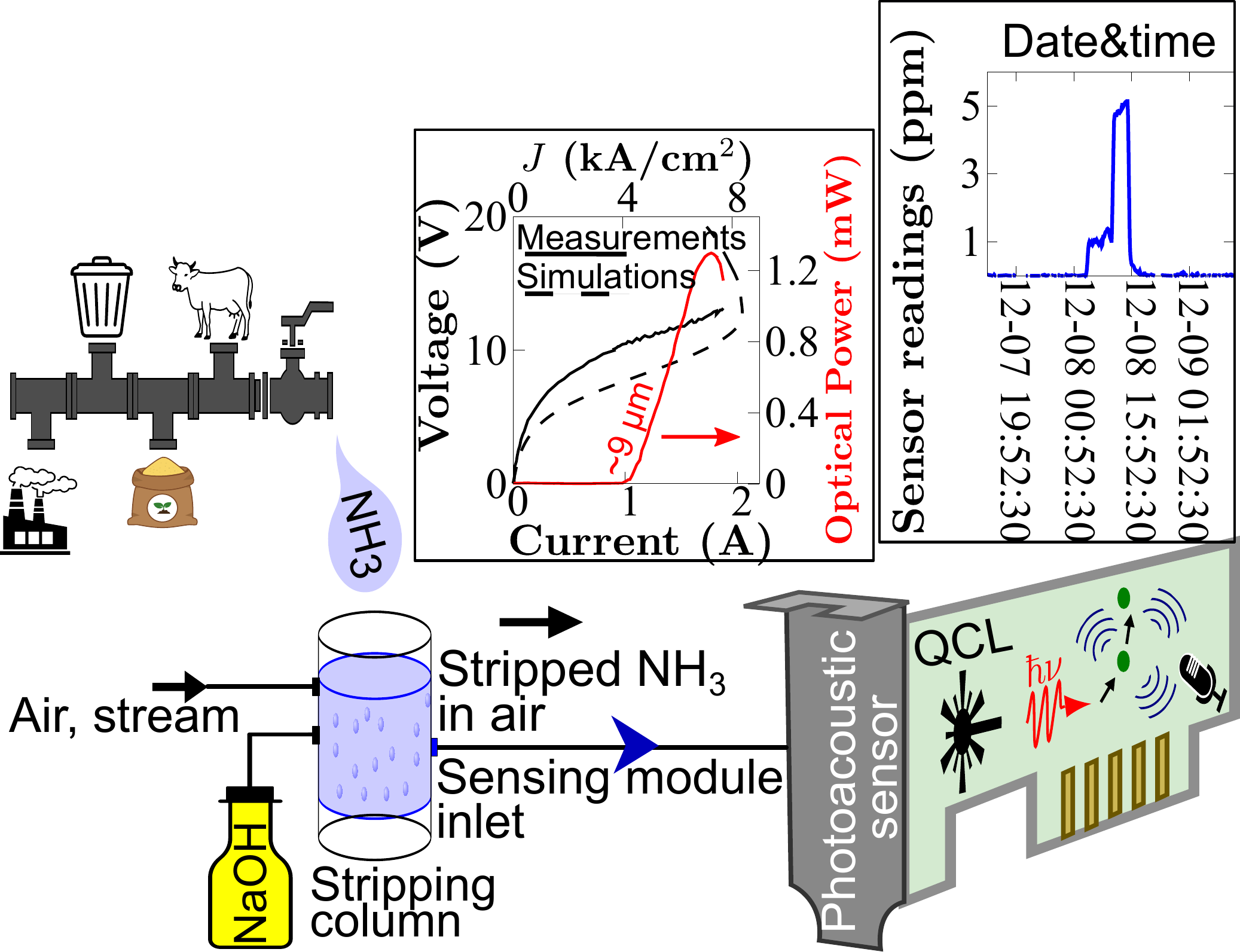}

\end{tocentry}

\twocolumn[
\begin{@twocolumnfalse}
\oldmaketitle
\begin{abstract}
\footnotesize
Ammonia (NH$_3$) toxicity, stemming from nitrification, can adversely affect aquatic life and influence the taste and odor of drinking water. This underscores the necessity for highly responsive and accurate sensors to continuously monitor NH$_3$ levels in water, especially in complex environments where reliable sensors have been lacking until this point. Herein, we detail the development of a sensor comprising a compact and selective analyzer with low gas consumption and a timely response, based on photoacoustic spectroscopy. This, combined with an automated liquid sampling system, enables the precise detection of ammonia traces in water. The sensor system incorporates a state-of-the art quantum cascade laser as the excitation source emitting at 9 \textmu m in resonance with the absorption line of NH$_3$ located at 1103.46 cm$^{-1}$. Our instrument demonstrated detection sensitivity at low ppm level for total ammonia nitrogen with response times less than 60 seconds.  For the sampling system, an ammonia stripping solution was designed resulting in a prompt full measurement cycle (6.35 mins). A  further evaluation of the sensor within a pilot study showed good reliability and agreement with the reference method for real water samples, confirming the potential of our NH$_3$ analyzer for water-quality monitoring applications.
\end{abstract}
\end{@twocolumnfalse}
]
Ammonia (NH$_3$) is a primary water pollutant present at varying concentrations in both groundwater and surface water. When this nitric waste reaches high levels in water, aquatic organisms find it challenging to discharge the toxicant effectively, leading to toxic buildup. This, in turn, affects the population dynamics of fisheries \cite{romano2013toxic}. Additionally, ammonia generated in sediments due to nitrification can be toxic to benthic organisms \cite{triest2001comparative} and surface water biota \cite{hora2020increased}.

The widespread use of ammonia on farms and in industrial or commercial locations indicates that exposure can occur due to accidental releases \cite{anjana2018toxic, schilt2004ammonia}. Moreover, the possibility of deliberate events, such as a terrorist attack involving ammonium nitrate \cite{makarovsky2008ammonia}, cannot be ignored. Ammonium nitrate is typically synthesized from nitric acid and household ammonia products \cite{michalski2015oxygen}.

Continuous water quality monitoring of ammonia is becoming increasingly important for plant operations and the quality control of water utility companies. Taste and odor problems, along with decreased disinfection efficiency, can arise if chlorinated drinking water contains more than 0.2 mg of ammonia per liter \cite{european2012health}. The drinking water standard recommended by the US National Academy of Sciences \cite{roney2004toxicological}, adopted by many European nations, is 0.5 mg/l (ppm).

As discussed above, uncontrolled releases of ammonia into the environment have a significant impact on human health, ecosystems, fiscal activities, and climate. This underscores the need for accurate and rapid detection techniques capable of assessing ammonia levels in water over a broad range of concentrations.
\begin{figure*}[!t]
   \includegraphics[width=0.85\linewidth]{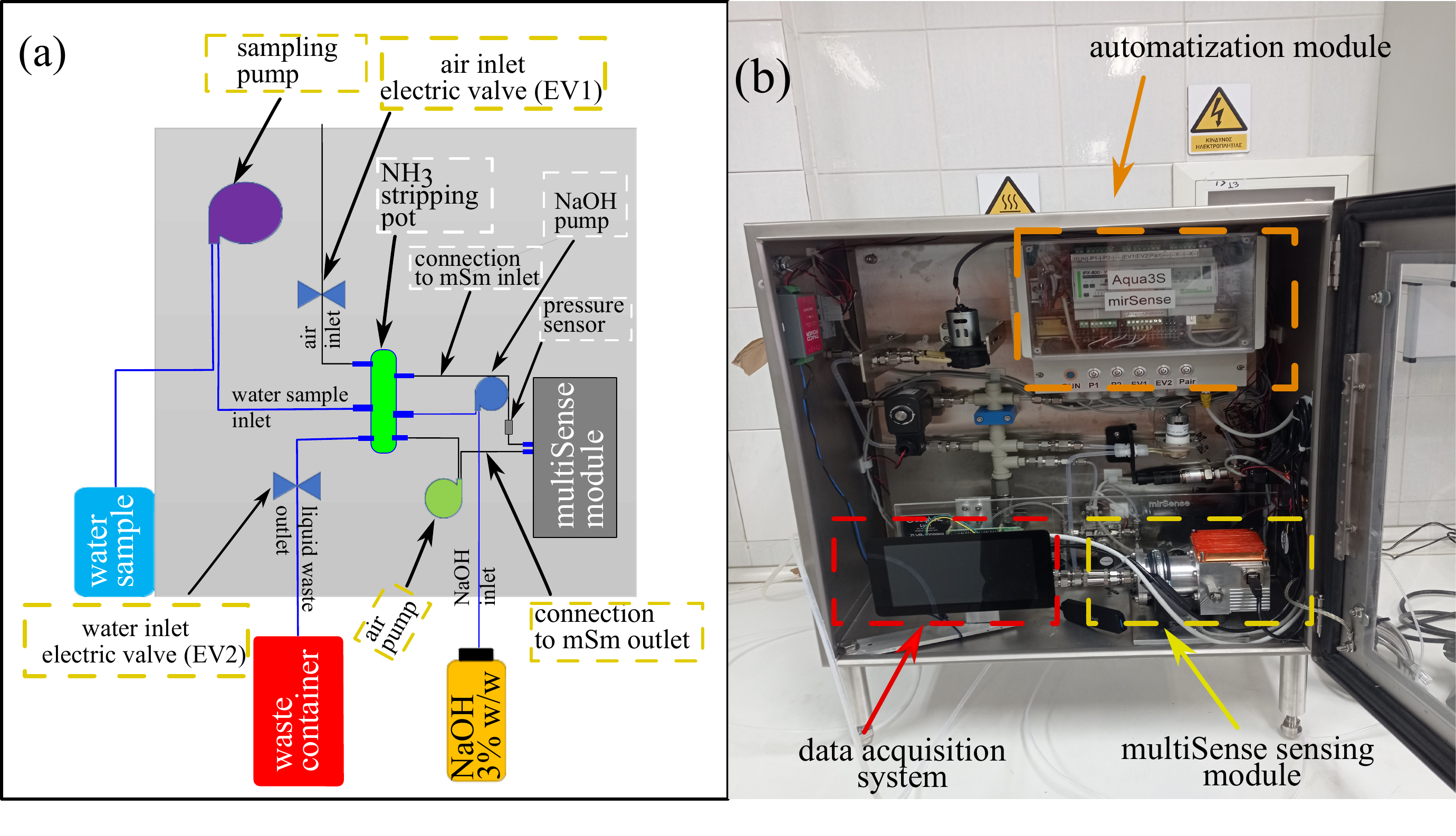}
   \caption{(a) Schematic setup of the gas and water sampling lines and flow in the ammonia analyzer. (b) A picture of sensor including the multiSense sensing module (mSm), the data acquisition system and the automatization box.}
\label{fig1}
\end{figure*}

Although sensor technologies exist for low ppm detection levels of ammonia, quality assurance and reliability for long-term sensing in complex environments is still lacking \cite{li2020detection}. Furthermore, some electrochemical sensors suffer from sensitivity to high background ion concentration
and the influence of actual field conditions (e.g. pH, humidity, salinity, temperature) \cite{cammerer2020application}. Before proceeding further, a brief comment on complementary Terahertz (THz) detection technologies  is due. Substances such as NH$_3$ and  deuterated ammonia (NH$_2$D)  have several absorption signatures in the THz range (100 GHz to 10 THz).  Superlattice multipliers   have been recently used to study nitriles in urine as potential markers for kidney damage after chemotherapy \cite{vaks2022sensing}. These type of devices take advantage of  nonlinear processes \cite{apostolakis2020superlattice}
 by acting as frequency multipliers \cite{pereira2021combined} of electromagnetic waves. While in the present work, NH$_3$ is the target, in Ref. \cite{vaks2022sensing}, the NH$_3$ resonances were so strong in a large spectral range, from to 140 GHz to 791 GHz, that they needed to be avoided. 
Recent advances in spectroscopy technologies in the mid-infrared (MIR) region \cite{jouy2014mid, haas2016advances} and    near-infrared (NIR) region \cite{pasquini2018near} have been exploited to enhance the performances of on-line water quality monitoring methods. Both approaches can reveal significant details about the molecular-level understanding and chemical properties of the water sample under study. However, spectra in the MIR region provide more specific information about the fundamental molecular vibrations (e.g. stretching, bending, scissoring, wagging), whereas absorption in the NIR region stem either from combination or overtones of fundamental vibrations \cite{manley2014near}.  Quantum Cascade Lasers (QCLs) operate in pulsed or continuous wave (CW) mode, at room temperature, with high output power and efficiency for MIR devices which have already demonstrated important applications in the field of trace gas detection using photo-acoustic (PA) spectroscopy \cite{cristescu2008laser, zeninari2020widely}, imaging \cite{yang2021phase}, and recently for molecules detection in aquatic solutions. Current paradigms involving QCL-based spectrometers suitable for water sensing include a thin-film waveguide flow cell accessory coupled to a broadly tunable QCL (between 10.52 and 11.23 \textmu m) facilitating low ppm detection ($\sim$ 5 ppm) of chlorinate hydrocarbons traces in water \cite{haas2016sensing}   and a chip-based evanescent wave sensing platform (QCL light emitted between 6.5 and 7.5 \textmu m) allowing the detection of aqueous toluene in a limit of 7 ppm \cite{beneitez2020mid}. \\ In spite of the certain progress 
in the laser based water sensing \cite{jouy2014mid, haas2016sensing,beneitez2020mid, freitag2020polarimetric}  and gas sensor applications \cite{yadav2021gas}  for water quality monitoring,  automated and reliable systems using these sensing techniques for detection of ammonia in water is far from being achieved \cite{goldshleger2018real}. 
\begin{figure*}[!t]
\includegraphics[width=0.9\linewidth]{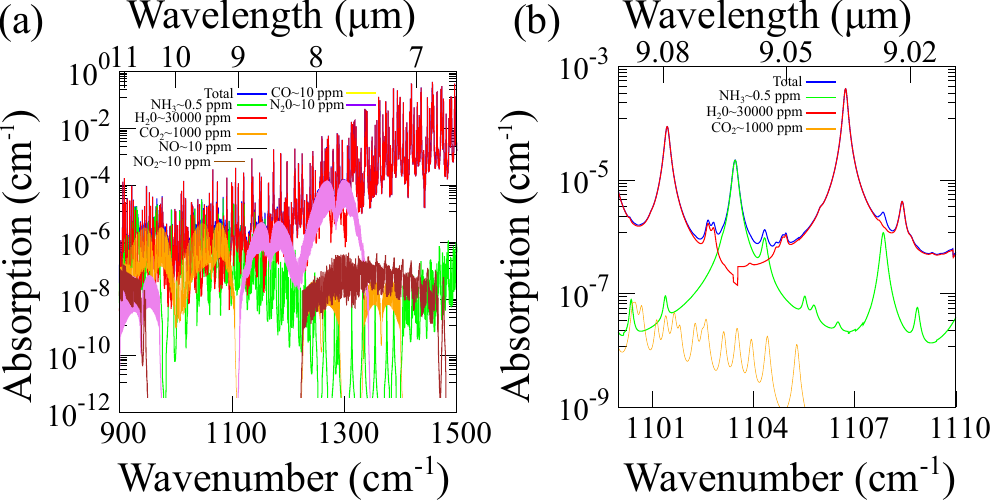}
   \caption{(a) Spectroscopic simulations of a gaseous matrix containing ammonia gaseous within the mid-infrared region. (b) The recommended absorption band for the ammonia detection in the spectral region from 1100 cm$^{-1}$  to
   1110 cm$^{-1}$. The temperature is considered $T$=333 K whereas the pressure is $P=$1 atm.}
\label{fig2}
\end{figure*} 
\\In this work we present a sensor based on a photoacoustic mid-infrared spectrometer trace gas analyzer and an automated ammonia stripping method to determine the levels of ammonia-nitrogen concentration in water samples. A photoacoustic gas sensor device  is a device
capable of analyzing gas based on the photoacoustic effect \cite{schilt2004ammonia,maurin2020first}.
In our system, the photoacoustic gas sensor device  incorporates an QCL emitter module  generating mid-infrared light pulses to
be absorbed by a gas containing ammonia molecules which can be stripped from the water phase into the gas phase by a specially designed stripping system. The non-radiative relaxation of the excited molecules results in an acoustic wave which is detected by a pressure-sensitive module and thereafter generate a corresponding sensor signal.  Detection limits are established both for the gas-analyzer and  the overall performance of the system for tracing ammonia molecules in water.  Moreover, the ammonia analyzer underwent sensitivity and stability tests to optimize the total measuring and sampling run cycle in laboratory conditions, in addition to real-time   quantitative analyses of ammonia under field conditions in realistic test cases. The results obtained indicate the potential of the sensor for near real-time water monitoring involved in the control of wastewater plants and environmental applications. 
\section{EXPERIMENTAL SECTION {\label{sec:level2}}}
\subsection{Chemicals and preparation of liquid samples. {\label{sec:sub1level2}}}
A standard solution of each water sample was prepared by following the standards \cite{o1993environmental}. Specifically, ammonia stock solution (1000 ppm) was prepared by  dissolving accurately 3.819 g of anhydrous ammonium chloride in deionized (DI) water and diluting to 1000 ml in a  volumetric flask, whereas the ammonia working standards (e.g. 
10 ppm) by diluting 10 ml of ammonia stock solution to 1 L of DI-water in a volumetric flask. The reagent  solution (0.8 ml of NaOH 3$\%$ w/w per measurement cycle) ensured the maximum removal efficiency of ammonia.
\subsection{Integrated sampling operation of \textnormal{NH$_3$} analyzer.}
Figure \ref{fig1}(a) shows the schematic diagram of the sampling lines in the ammonia analyzer which allows the automated water sampling.  The assembly of the ammonia stripping/liquid sampling sub-system was provided by Swagelok \cite{swagelok} following our design and technical specifications. After preparing the reagent and water samples as discussed in the previous \hyperref[sec:sub1level2]{subsection},
they were loaded into the corresponding containers. First, EV1 valve opens allowing air flow while the air-pump (diaphragm pump KNF-NMP05) is powered on in order to bring fresh air at a mean rate of 450 ml/min  from outside the enclosure   into the sampling circuit. Thereafter, the water sample is injected into the stripping column using a peristaltic sampling pump at a flow rate of 220 ml/min and subsequently a different peristaltic pump  is turned on to add NaOH solution at a flow rate of 28 ml/min into the same reservoir. Thereafter, EV1, EV2 valves are closed whereas   the air pump is re-activated  to  flush air within the column igniting the stripping process of NH$_3$ and thereafter  the tail gas, a mixture of air and ammonia, is released through the top of the stripping column and then absorbed by the PA gas sensing module.  Finally, by reopening  the EV1, EV2 valves results in transferring the analyzed water sample into the waste container and therefore completing the full-sampling cycle. To fill the stripping column with water sample, a silicone tubing of 1.5 m length$\times$6 mm outside diameter (O.D.), was connected to the sampling pump using a tubing adapter. The  same type of silicone tubing was used to connect the  outlet of the stripping pot to the waste container, whereas a PFA tubing (1.5 m length$\times$3.18 mm O.D.) was connected to the peristaltic pump to inject the reagent solution to the striping pot.
\subsection{Spectral range selection process.}
The measurement method relies on stripping of free ammonia (Figure \ref{fig1}a) from water into the gas phase where the spectrometer module (Figure \ref{fig1}b)  can effectively detect the target molecules. Thus, it is critical to identify the wavelength corresponding to the light of the laser source which can be absorbed by a specific gaseous specie. In our case, the N-H bond wagging vibration mode in the proximity of 8.78 \textmu m \cite{stewart1999surface}  has been chosen  due to strong fundamental absorption of  the ammonia molecule in the spectral region  $8.6-9.1$ \textmu m,  which can be covered by the lasing technology included in the MIR sensing module, whereas the other MIR regions  reveal pronounced overlaps with
 absorption bands of H$_2$O, and therefore, they are not suitable for  detection of stripped ammonia. The selection of this spectral window was also proposed by  Owen et al. \cite{owen2013measurements}.   The simulated absorption spectra in Figure \ref{fig2}a of the gas  matrix  including simple gas molecules such as water (H$_2$O), carbon dioxide (CO$_2$), Nitric Oxide (NO) and the targeted gaseous ammonia, was performed using the parameters from HITRAN database \cite{hitran2023}. Therefore, the carrier-gas stream involved in the stripping process  is considered of typical composition as found in moist air \cite{picard2008revised}.  To ensure that the ammonia enriched gas matrix can be modeled realistically, the following features was assumed: (i) the gas molecule of interest is present at a significantly low concentration ($\sim$ 0.5 ppm), (ii) the other interfering species, which are carried by the carrier-gas are present at their maximum potential concentration, e.g. levels of CO$_2$ and H$_2$O at $\sim$1000 ppm and  30000 ppm respectively. The difficulty of measuring ammonia is further depicted in Figure \ref{fig2}b. This zoom-in plot of the absorption spectra highlights a potential peak interferent due to water vapor which can be circumvented by choosing a wavelength near 1103.46 cm$^{-1}$ corresponding to a maximum absorption intensity $\alpha_{G,max}=2.23$ $\times$10$^{-5}$ cm$^{-1}$, where the absorption lines of CO$_2$ and H$_2$O in particular is weaker. 
 \subsection{Photo-acoustic spectrometer design.}
The  MIR spectrometer module (multiSense) developed by mirSense company \cite{mirSense} was employed, which consists of a QCL laser  and a photoacoustic cell allowing detection of gas molecules at sub-ppm limit. In detail, the multiSense  module (mSm) has been fabricated as a semi “stand-alone” solution ($170\textrm{mm}\times 110\textrm{mm} \times 110\textrm{mm}$)  within a rack of $W550\textrm{mm}\times H575\textrm{mm} \times D300\textrm{mm}$, where adequate sampling conditions are
assured, i.e. heated sample lines, flow control, pressure measurement and software control (Figure \ref{fig1}b). The gas mixture retrieved by the stripping process is considered an air mixture as indicated in the previous section with a temperature ranging between 263 and 323 K and a typical pressure of 1 atm. Thus, the measurement gas cell within mSm is regulated at 333 K well beyond the upper limit of 323 K to make certain that the cell temperature will not drift due to an elevated temperature. The multiSense module employs the conventional photoacoustic spectroscopy (PAS) technique  using commercial microphones.  The QCL operates in pulsed mode, allowing to reduce the heat produced within the system \cite{carras2023photoacoustic}.
\begin{figure}[!t]
\includegraphics[width=8.1cm]{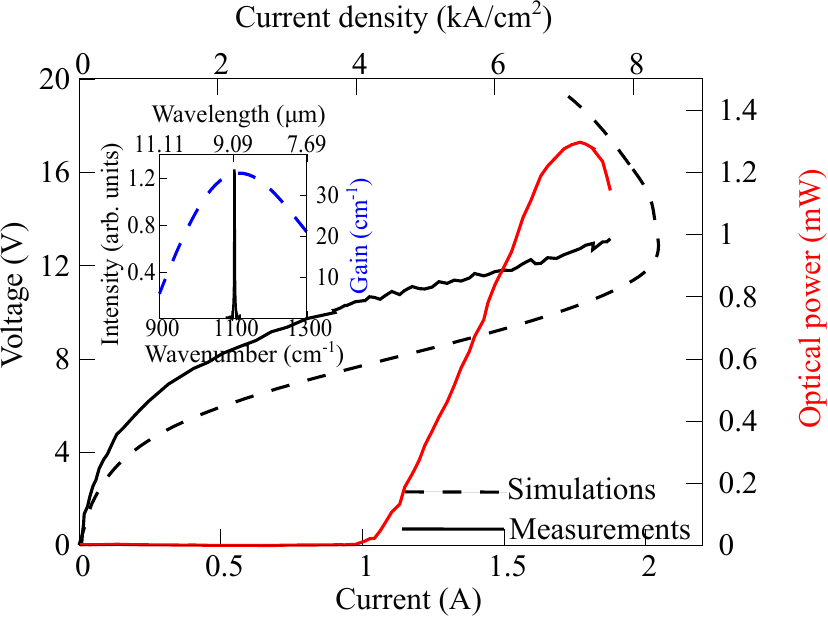}
   \caption{  Power-Current-Voltage (PIV) characteristics of a QCL structure
InGaAs/AlInAs
at room temperature. The dashed and solid
lines depict the simulated curve calculated with a NEGF approach
and the measured characteristics respectively.
The inset illustrates the obtained optical spectra in comparison to the numerical calculations of the optical gain.}
\label{fig3}
\end{figure} 
In particular, the QCL source emits radiation 
close to selected absorption line ($k\sim$ $k_{\mathrm{NH_3}}$) illustrated in Figure \ref{fig1}b, while operating in a quasi-continuous wave regime at room temperature. The design of the QCL structure is based on InP wafer, whereas the repeats of AlInAs/GaInAs ternary layers correspond a double-phonon resonant design with four quantum
well active region optimized at $\lambda \sim$ 9 \textmu m. The simulated current-voltage  characteristic based on a  Nonequilibrium Green's Functions (NEGF) method  \cite{Tim,Winge_2016, pereira2017theory, pereira1998microscopic} agrees well with the experimental one as shown in Figure \ref{fig3}. The voltage difference of experiment to simulation  is attributed to  to the non-optimized input  parameters of the scattering self-energies of every implemented scattering mechanism \cite{Winge_2016}.  The inset also demonstrates  a good fit between the
calculated peak gain frequency and the measured frequency of the  laser. 
The corresponding optical spectrum of the laser obtained with 300 ns long pulses and with a duty cycle of 3 $\%$ and a temperature of 293 K. These operational parameters were used for the electrical characterization of the QCL structure and they do not represent the laser working parameters during the process of the photoacoustic detection.   To achieve single frequency emission, a Bragg grating was used to select the wavelength. The period of grating allows to tune the emitted wavelength over the gain of the active region. Furthermore, the laser is a double trench design and has been epi-up mounted on the Aluminum Nitride submount.
\\  To determine the minimum optical power $P_l$ required by the emission of the fabricated QCL, we started by assuming a  normalized noise equivalent absorption (NNEA) coefficient of $2\times 10^{-8}$W cm$^{-1}$Hz$^{-1/2}$ which is a known metric of the photoacoustic detector's sensitivity \cite{sampaolo2016improved, maurin2020first} and a detection bandwidth $\Delta f=1.7 \times 10^{-2}$ Hz  bandwidth (inversely proportional to detection time $\Delta T=60 $ s). The minimum detectable absorption is commonly defined as $\alpha_{G,min} (3\sigma)=\textrm{NNEA} \sqrt{\Delta f}/P_l$ \cite{baudelet2014laser}.  Thus, given the selection of the absorption peak $\alpha_{G,max}$ illustrated in Figure \ref{fig2}, the required optical power is $P_l=\textrm{NNEA} \sqrt{\Delta f}/\alpha_{G,max}\sim 0.12$ mW and thereafter we adjusted accordingly the duty cycle  to match this optical power. The overall parameters adjustments of the PA spectrometer resulted in a detection limit ($3 \sigma$) of 0.18 ppm for gas-NH$_3$ concentration in 60 s averaging.
 In addition, the linearity of measurement within
a 0$-$500 ppm range was demonstrated. Resolution below 0.1 ppm was also shown for gas phase
measurements with mSm module. For further details of the mSm module operation refer to Figures S1-3 of \hyperref[sec:suppinfo]{Supportive information}.\\
The  features of the photo-acoustic spectrometer as presented here indicate its potential  for determination of traces of ammonia in water by means of NH$_{3}$ stripping which allow the separation of free-ammonia into the gas phase.
 \begin{figure}[t]
   \includegraphics[width=0.95\linewidth]{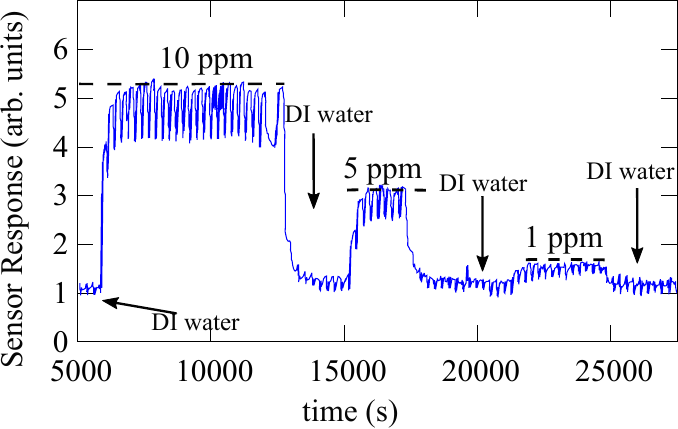}
   \caption{Dynamic response of the sensor to three different concentrations of NH$_3$ in water at room temperature. These records are interrupted by deionized water rinse.}
\label{fig4}
\end{figure}  
\subsection{Ammonia stripping process.}
The ammonia stripping process is a stripping method based on the principle of mass transfer \cite{matter1981transfer,kim2021influence}. The ammonia stripping/sampling compartment of the analyzer was structured as depicted in Figure \ref{fig1}a, with the stripping column positioned at the center of the rack. Specifically, in this method water is contacted with air to strip the ammonia gas present in the water. The presence of ammonia in water can be found in two forms, namely, ammonium ions (NH$_4^+$) and ammonia gas (NH$_3$). The relative concentrations of ammonia gas and ammonium ions are directly dependent on the pH and the temperature of water. 
Ammonia nitrogen in water exists in equilibrium between the molecular and ionic form according to the following reaction
\begin{equation}
\mathrm{NH_3+H_2O \leftrightarrow NH_4^+ + OH^-},
\end{equation}
whereas the dissociation of water is given by the equilibrium reaction
\begin{equation}
\mathrm{H_2O \leftrightarrow H_3O^+ + OH^-.}
\end{equation}
The ammonia fraction, $f_{\mathrm{NH_3}}$, determines the concentration ratio between free ammonia $[\mathrm{NH_3}]$  and total ammonia $[\mathrm{NH_3}]^{total}=[\mathrm{NH_3}]+[\mathrm{NH_4^+}]$.  Typical stripping processes require a sample temperature between 293 and 323 K, whereas pH values range between 10 and 12. This follows from the dependence of free ammonia on pH and temperature (see Figure S4). Therefore, a basic solution acting as reagent  is needed to regulate the pH levels. In our case,  a NaOH solution was used, resulting in significantly alkalinized pH$\sim12$ and therefore to an ammonia fraction close to one.  Complementary description of the ammonia stripping process is given in \hyperref[sec:suppinfo]{Supportive information}.

 \section{RESULTS AND DISCUSSION}
\begin{figure}[t!]
   \includegraphics[width=7cm]{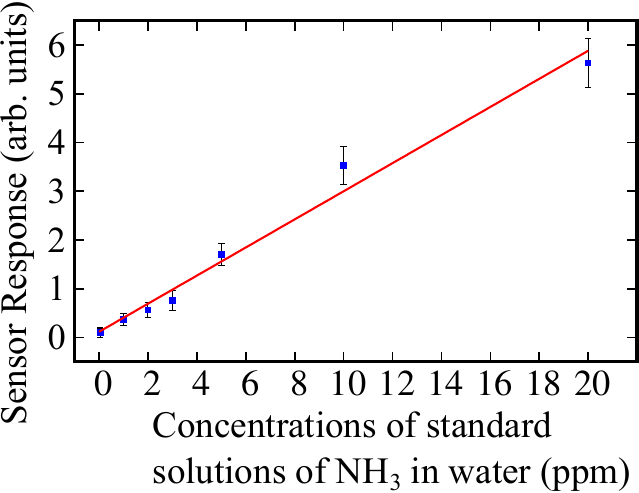}
   \caption{Sensor response (calibration curve) within the range of 1 to 20 ppm of NH$_3$ concentration in water at room temperature. }
\label{fig5}
\end{figure} 
The sensor calibration for the detection of ammonia traces in water was achieved by using standard reference solutions (see \hyperref[sec:sub1level2]{Chemicals and preparation of liquid samples}) starting from a NH$_3$ stacking solution of 1000 ppm which is diluted to obtain reference samples of lower concentration of ammonia. The monitoring of ammonia is hindered by the sticky nature of its polar molecule \cite{schmohl2001effects} that commonly adheres to inert surfaces. To prevent the attenuation of sticky molecules such as ammonia within the tubes during the sampling process, we carefully storage the samples under refrigeration. On that account, samples with above 233 K were avoided in order to suppress the condensation risk in the pipelines.
\begin{figure}[t!]
   \includegraphics[width=0.8\linewidth]{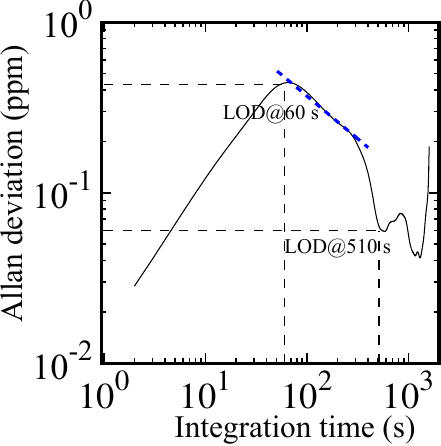}
   \caption{Allan-Werle deviation analysis recorded on 10 ppm of ammonia concentration in water.Blue dashed line indicates a typical  dependence ($1\sqrt{t}$) for white noise.}
\label{fig6}
\end{figure} As is  well known,   condensation makes NH$_3$ to be dissolved in water droplets immediately \cite{dasgupta1986solubility},
compromising the reliability of the measurement. Figure \ref{fig4} demonstrates consecutive records with reference samples 10 ppm, 5 ppm and 1 ppm of ammonia concentration. Each oscillation represents a measuring cycle (6.35 min corresponding to 1 s time step) including the sampling time of the water sample $\sim$ 0.4 min, the sampling time of  the NaOH reagent solution $\sim$ 0.05 min and the purge time $\sim$ 0.7 min. These records are intermitted (vertical arrows)  by measurements of blank solutions; 2 ml NaOH-DI water samples were used as  blank solutions which indicated that their related memory effects do  not affect significantly the consecutive measurements. In turn and as anticipated, the relationship between the reference samples and the sensor response (Figure \ref{fig5}) is linear (Pearson's $r=0.99$). Note that by considering a low time constant and in 60 s of averaging, we obtain a LOD just about 0.5 ppm. By resorting to an Allan-Werle variance method \cite{allan1966statistics} with time constant of 1 s and a 10 ppm concentration of ammonia in water, the limit of detection may reach down to 0.4 ppm  for integration time of 60 s (Figure \ref{fig6}). With a  increase of averaging time, the Allan-Werle plot demonstrates a continuous decrease,  revealing a white-noise prevalent behavior ($\propto t^{-1/2}$)  of the sensor response as shown by the blue dashed curve. Moreover,  Figure \ref{fig6} suggests that the minimum limit of detection  can be further improved to 0.06 ppm with an integration
time of 500 s, whereas longer integration time clearly indicates the emergence of a long-term drift.  

\subsection{Photoacoustic NH$_{3}$ analyzer: pilot studies.}
\begin{figure}[t!]
   \includegraphics[width=0.85\linewidth]{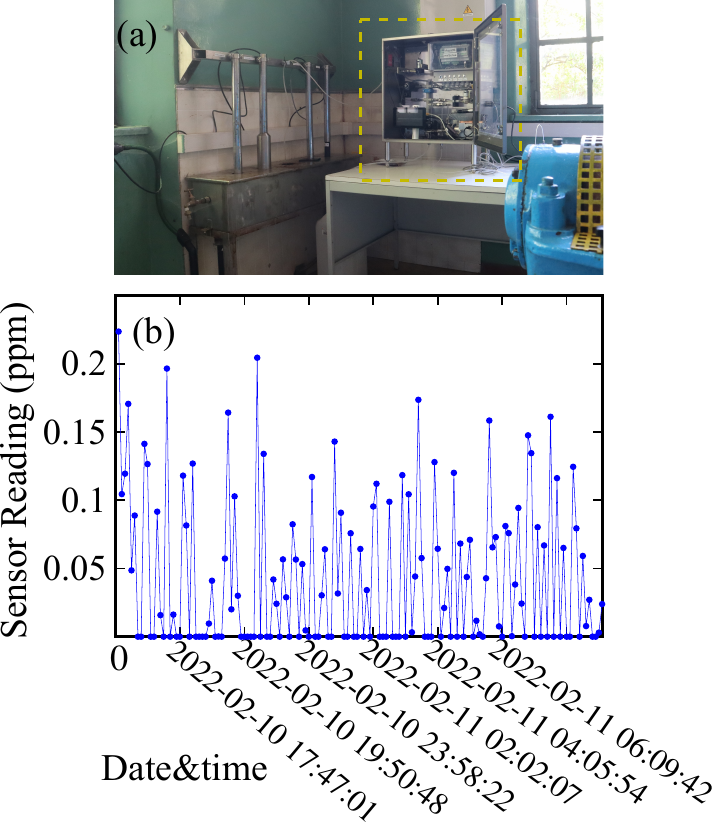}
   \caption{(a) Field test of the sensor (highlighted by the yellow frame) which was deployed in a pumping station  during a pilot case study. (b) Sensor Reading as a function of time indicating the concentration of ammonia traces within spring water.}
\label{fig7}
\end{figure} 
The laboratory results indicated that the designed device can
effectively detect the presence of ammonia in water. To validate and further develop this innovative technology, the sensor was installed in different pilot cases within the context of H2020 project aqua3S \cite{aqua3S2023} which aimed to standardize existing sensor technologies complemented by state-of-the-art detection mechanisms focusing on water safety and security. 
In particular, the sensor was deployed at Trieste Aqueduct (Randaccio site), Italy  and Quality Control laboratory located at
the Thessaloniki Water Treatment plant (TWTP), Greece. In the first pilot case, the sensor was continuously monitoring the water stored in a pumping station (Figure \ref{fig7}a), showing practically no response (Figure \ref{fig7}b) which is consistent with the historical data ($<0.05
$ ppm) of ammonia concentration determined by analytical measurements; the currently estimated LOD of the sensor is above this range.
\begin{table*}[h]
\centering
\caption{Recovery study performed by adding standard solutions of ammonia to real water samples}
\begin{tabular}[t]{cccccc}
\hline  NH$_{3}$ added (ppm)& \thead{NH$_{3}$ found \\ before spiking (ppm)}& Expected (ppm) &
& Measured (ppm) & Recovery ($\%$)\\
\hline
0.5&0.08&0.58&&0.72&124.4\\
1&0.08&1.08&&0.72&147.2\\
2&0.08&2.08&&2.59&124.4\\
3&0.08&3.08&&3.5&113.6\\
5&0.08&5.08&&5.69&112\\
\hline
\end{tabular}
\label{table}
\end{table*}%
\begin{figure}[ht!]
   \includegraphics[width=0.9\linewidth]{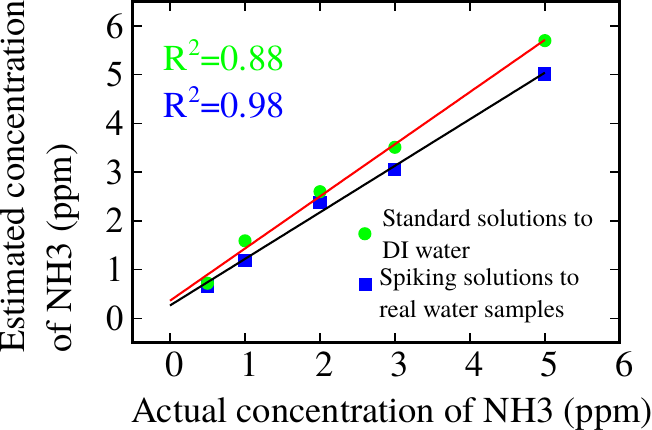}
   \caption{Determination of ammonia concentration in real water samples (black curve) and deionized water (red curve).}
\label{fig8}
\end{figure} 
In the second pilot case, the optimized sensor system which is equipped with automated sampling, was applied first for determination of  NH$_3$ in real samples collected from the $\Delta 2$ tank of the TWTP, spiking spiking different ammonia concentrations directly into the water samples.   
Figure \ref{fig8} displays a good linear response (black curve) of the sensor to  real water samples with with a linearly
dependent coefficient $R^2=0.98$. 
This result indicated the reliability of the approach with practically no-matrix  effects due to formation un-ionized ammonia at high pH values resulting from the use of NaOH-buffer solution. The wide linear range of the sensor  comfortably allowed measurements for ammonia concentration larger than 2 ppm which were critical for the implementation of pilot scenario considering an extended ammonia spillage event in the water storage tank of TWTP. The recovery values of ammonia
are presented in Table \ref{table}, showing a recovery range from 112 to 147.2 $\%$. Furthermore, the  enhanced recovery rates  can be attributed  to 
on-going conditioning or measuring cycles which might resulted in minor memory effects due to residual ammonia molecules released in the course of successive
measurements. 

 \section{CONCLUSIONS AND OUTLOOK}
The work presented here describes the development and the characteristics of a sensor system suitable for detection of ammonia traces in water by employing a photoacoustic sensing technique.  To the best of our knowledge, this is the first demonstration of a QCL-based ammonia detector combined with an automated ammonium ions  stripping process. The NH$_3$ analyzer demonstrated high sensitivity with a  detection limit  of 0.5 ppm, which is considerably low once compared with other devices incorporating QCL structures for detection of molecules in water or electrochemical gas sensors integrated with ammonia stripping modules. Real-time monitoring of ammonia concentration in real water samples were performed in the course of pilot field case studies and the sensor could detect ammonia in real water samples. As a result, these results outline the potential of gas sensor applications and photoacoustic sensing particularly in the  systematic water quality monitoring. Currently, we design anew our water sampling method to allow detection of other molecules in water beyond ammonia. In order to extend the feasibility of our approach at part per billion levels, improvements in the water/gas sampling circuits should be considered such as reduction of the noise generated by  the air-pump and sampling pump, decrease of the cold points of the instruments in order to avoid water condensation issues, enhancement of the efficiency of the purging steps, redesign of the water reservoir to allow an increase of ammonia recovery at low concentration levels and finally further optimization of the QCL stucture and lasing power which can emit at the characteristic wavelength which is resonant with the vibrational excitation of ammonia molecules.
 \section{Author contributions}
 M.F.P. conceived the initial idea; G.A. and A.A. developed and refined the concept. M.F.P. and A.A. developed the theory, programmed the equations and analyzed the data. G.A., G.M., A.A. and L.L.  designed and implemented the experiments. All co-authors contributed to the text.

 \section{ACKNOWLEDGMENTS}
 This work was supported by the EU H2020$-$Europe’s resilience to crises and disasters program (no. 832876, aqua3S) and  EU H2020$-$Secure societies$-$Protecting freedom and security of Europe and its citizens (no. 101021857, Odysseus). M.F.P. further acknowledges support from Khalifa University of Science and Technology under Award No. CIRA-2021-108.  During the preparation of this work the authors used Chat-GPT in order to assist the process of manuscript proof-reading  after initial drafting. Furthermore, Turnitin, which detects plagiarism, has been used to evaluate the final text. After using this tools, the authors reviewed and edited the content as needed and take full responsibility for the content of the publication. 

\section{Supporting Information Available
{\label{sec:suppinfo}}}
Information is presented with respect to the processes involved in evaluating the performance of the gas sensor (mSm) response; Assessing the performance of the mSm in measuring gaseous ammonia and data on linear response range; data on sensor's response near the limit of detection.
Additional information regarding the ammonia stripping process under consideration is provided here.

\bibliography{biblio}

\end{document}


\newpage
\section*{Section S1}
Here, we present a set of figures (S.1-3) related to the processes involved in evaluating the performance of the gas sensor (multiSense module) response, along with corresponding data.
 \begin{figure}[!h]
\includegraphics[width=0.9\linewidth]{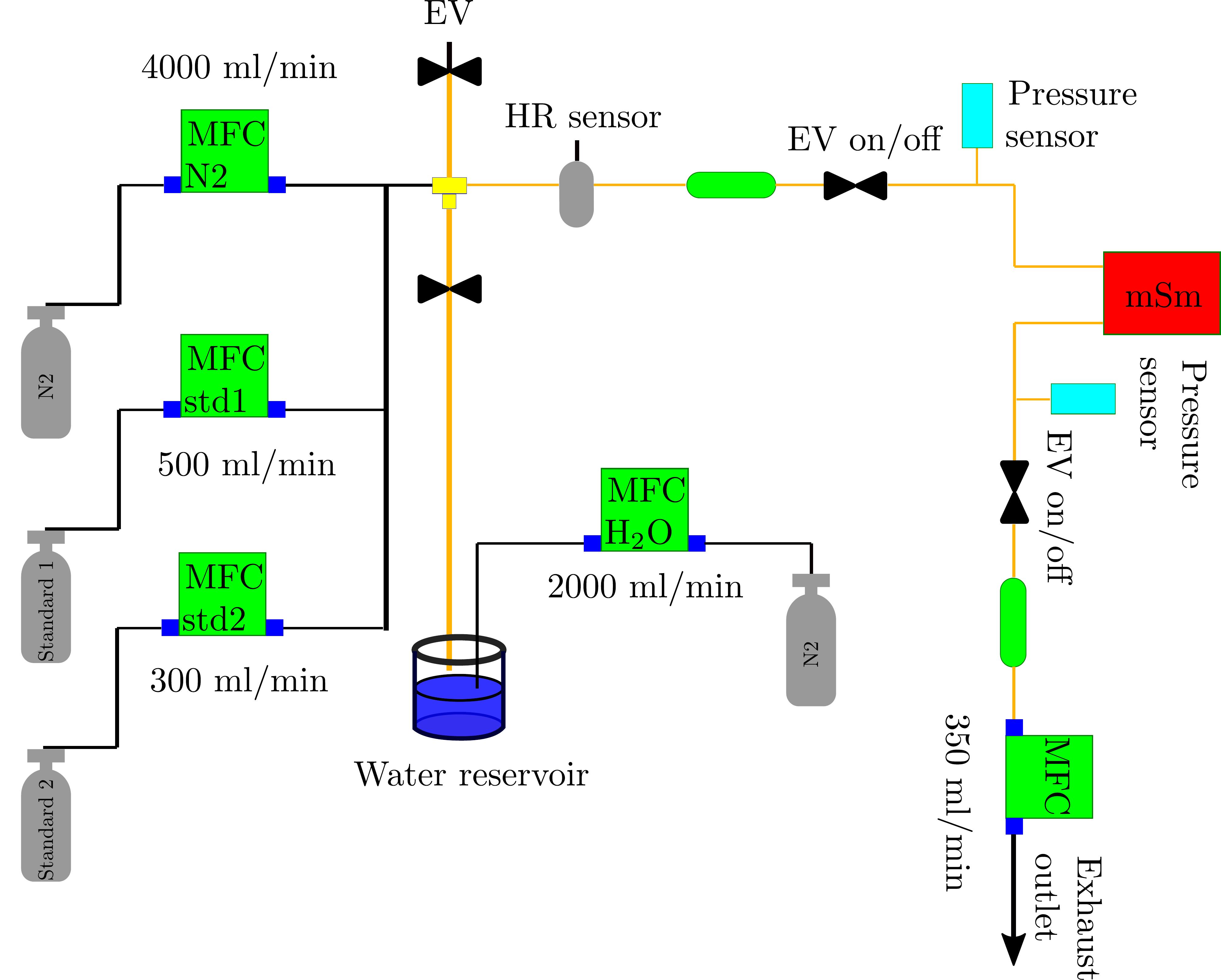}
   \caption{ \textbf{Simplified diagram of the laboratory test bench to evaluate the performance of the gas sensor response, i.e. multiSense module (mSm)}.  Different ammonia (NH$_3$) concentrations were sent to mSm by controlling the flow of the reference gas mixture by means of mass flow controllers, W (MFC) (Bronkhorst, Germany), which sets the final flow rate at 350 ml/min. This setup allows us to prepare various gas mixtures of high precision, providing several dilutions of NH$_3$ and nitrogen (N$_2$) calibrated mixtures. The pre-diluted test-gas is specified by analyte concentration of 500 ppm  NH$_3$ within a pressurized cylinder, whereas the N$_2$ mixture is introduced into the multiSense module after passing through humidifier stabilizing the temperature and the gas flow. All measurements are conducted under atmospheric pressure conditions, while the pressure itself is continuously monitored using pressure sensors.}
\label{figS1}
\end{figure}
 
 \begin{figure}[!h]
\includegraphics[width=0.9\linewidth]{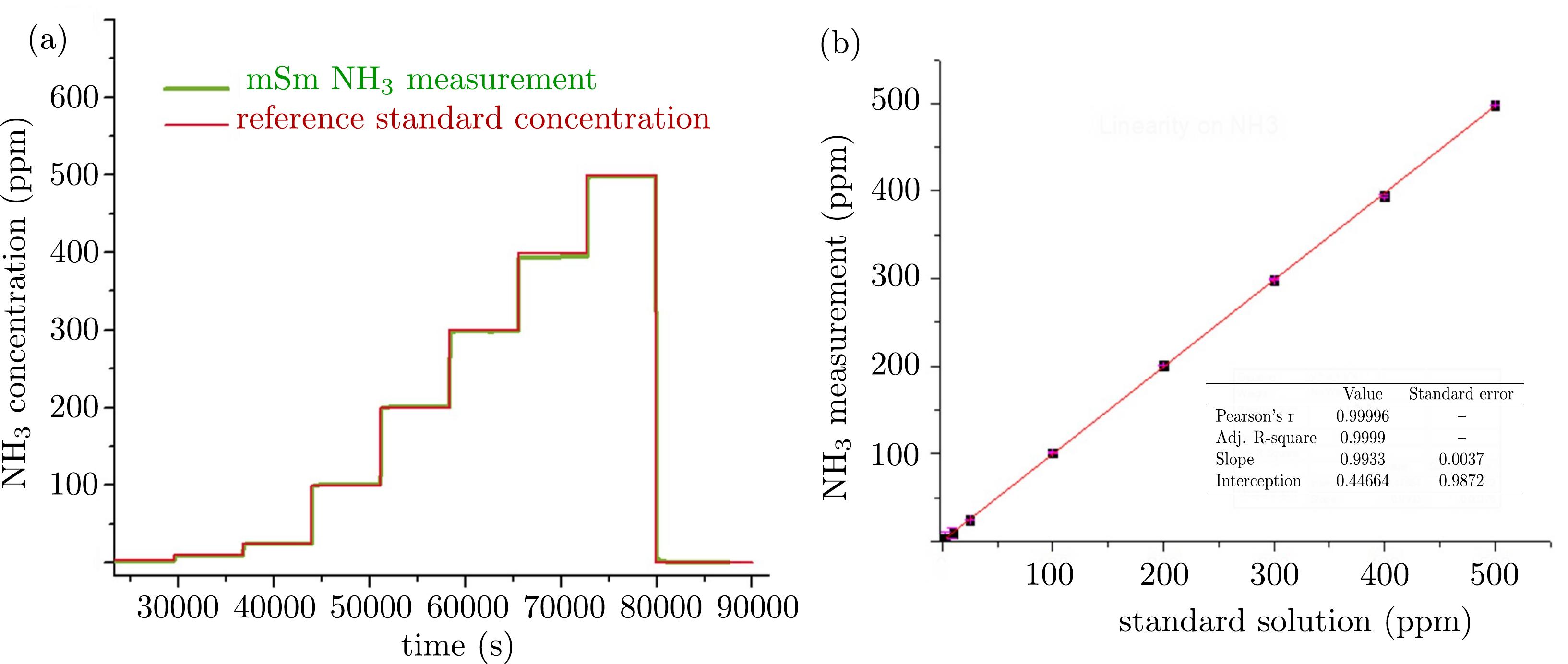}
   \caption{ \textbf{Assessing the performance of the mSm in measuring gaseous ammonia.} (a) A exhaustive data acquisition spanning 16 hours was conducted to evaluate the module’s linear response across the ammonia concentration range 0 to 500 ppm. (b) The linear relationship between the sensor’s response and the concentration of the standard solution samples. The inset contains a table of statistical information related to the linear response of the mSm. The mSm was securely housed within the analyzer rack to assess its thermal performance, electronic noise, and vibrations. A measured LOD of 0.18ppm in 60 s averaging was obtained.  Additionally, a resolution below 0.1 ppm was demonstrated for gas phase measurements using the mSm.}
\label{figS2}
\end{figure}

 \begin{figure}[!t]
\includegraphics[width=0.9\linewidth]{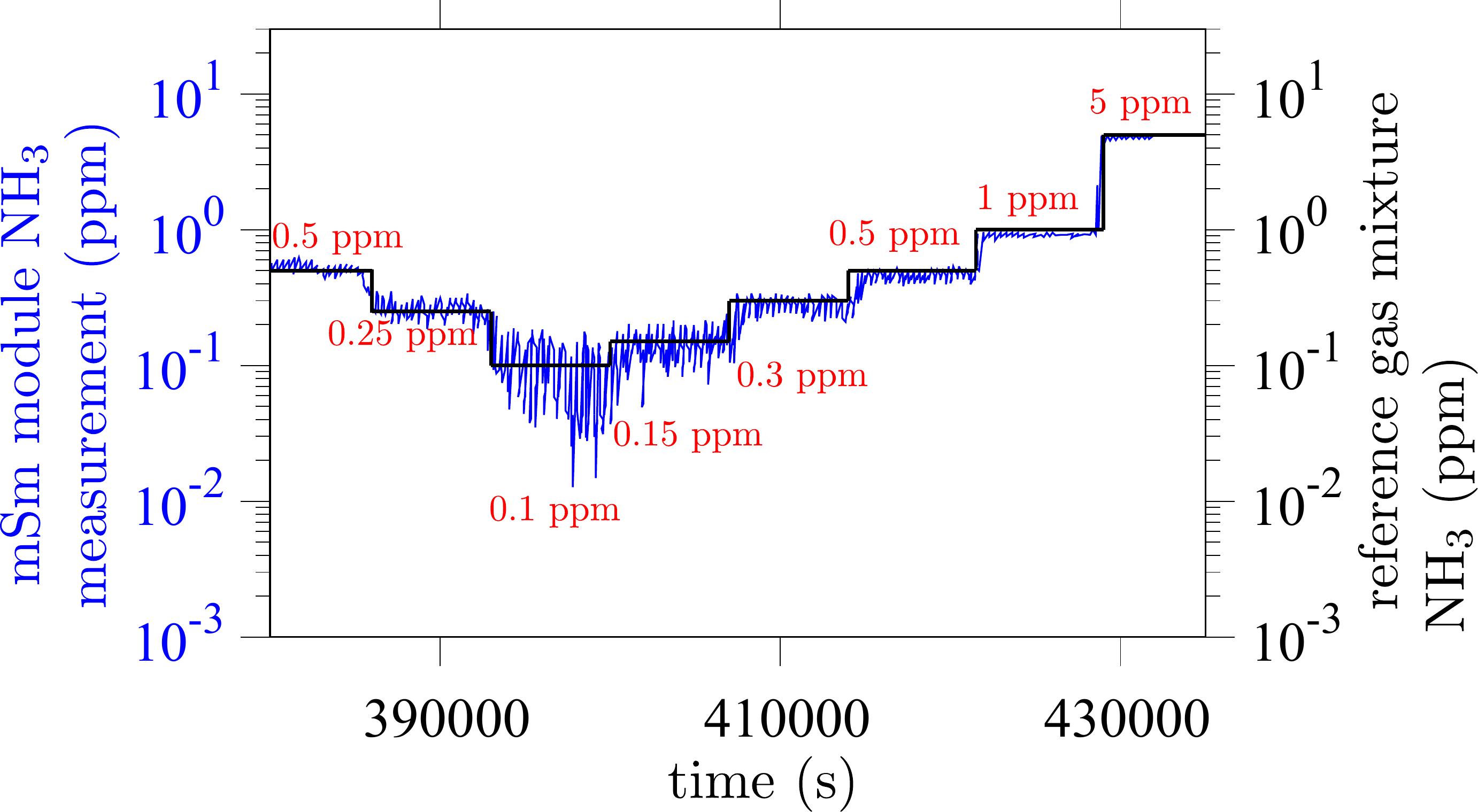}
   \caption{\textbf{Demonstrating the mS module's capability to measure traces of gaseous ammonia near the limit of detection.} The measurements are reproducible, and no hysteresis was observed. Each concentration step is maintained for two hours. This graph also illustrates a linear measurement range between 0 and 5 ppm of NH$_3$.}
\label{figS3}
\end{figure}
\clearpage

\section*{Section S2}
\begin{figure}[!t]
\includegraphics[width=0.8\linewidth]{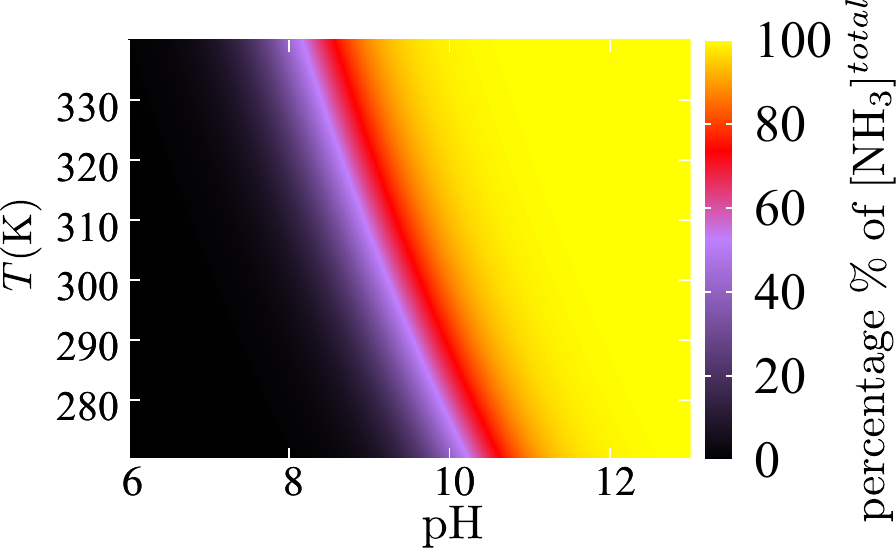}
   \caption{Colormap illustrating the dependence of free ammonia percentage [Eq. (\ref{eq:labelammstrip})] on pH and temperature.}
\label{figS4}
\end{figure}
In this section, we provide a more detailed description of the ammonia stripping method employed to extract gaseous ammonia from the water samples. The ammonia stripping process employs a mass transfer principle \cite{matter1981transfer,kim2021influence}. In this approach, water comes into contact with air, facilitating the removal of ammonia gas from the water. Ammonia in water exists in two forms: ammonium ions (NH$_4^+$) and ammonia gas (NH$_3$). The proportions of ammonia gas and ammonium ions are immediately affected by the water's pH and temperature. Ammonia nitrogen within water exists in a state of equilibrium between its molecular and ionic forms, as depicted by the following reaction
\begin{equation}
\mathrm{NH_3+H_2O \leftrightarrow NH_4^+ + OH^-},
\end{equation}
while the dissociation of water is described by the equilibrium reaction
\begin{equation}
\mathrm{H_2O \leftrightarrow H_3O^+ + OH^-.}
\end{equation}
The ammonia fraction, denoted as $f_{\mathrm{NH_3}}$, governs the concentration ratio of free ammonia $[\mathrm{NH_3}]$  and total ammonia $[\mathrm{NH_3}]^{total}=[\mathrm{NH_3}]+[\mathrm{NH_4^+}]$. It can also be defined as
\begin{equation}
f_{\mathrm{NH_3}}=\frac{1}{1+\frac{K_{\mathrm{NH_3}}}{K_{\mathrm{H_20}}}\times 10^{\mathrm{-pH}}},
\end{equation}
where
\begin{equation}
K_{\mathrm{H_20}}=[\mathrm{H_3O^+}][\mathrm{OH^-}]
\end{equation}
and 
\begin{equation}
K_{\mathrm{NH_3}}=\frac{[\mathrm{NH_4^+}][\mathrm{OH^-}]}{[\mathrm{NH_3}]}
\end{equation}
 determine the ionization constants of water and ammonia respectively. The ionization constants \cite{olofsson1975thermodynamics,bonmati2003air} and the ammonia fraction \cite{hansen1998anaerobic} have been calculated semi-empirically as
\begin{equation}
\begin{split}
\textrm{p}K_{\mathrm{NH_3}}=4\cdot 10^{-8}\cdot T^3(\textrm{K})+9\cdot 10^{-5}\cdot T^2(\textrm{K}) \\  -0.0356\cdot T(K)+10.072, \end{split}
\end{equation}

\begin{multline}
\textrm{p}K_{\mathrm{H_2O}}=24691.6\cdot T^{-1}(\textrm{K})+405.867\cdot \textrm{log}(T(\textrm{K}))     \\-0.488054\cdot T(\textrm{K})\\+2.37842\cdot 10^{-4}\cdot T^2(\textrm{K})-948.73, 
\end{multline}
and
 \begin{equation}
     f_{\mathrm{NH_3}}=\left(1+\frac{10^{-\mathrm{pH}}}{10^{-\left(0.09018+\frac{2729.92}{T(\textrm{K})}\right)}}\right)^{-1},
 \label{eq:labelammstrip}
 \end{equation}
 where $T(\textrm{K})$ is temperature in Kelvin. Typical stripping processes require a sample temperature between 293 and 323 K, whereas pH values range between 10 and 12. This stems from the dependence of free ammonia on pH and temperature as demonstrated in Figure \ref{figS4}. Hence, the requirement for a reagent in the form of a basic solution to control pH levels. In our experiment, we utilized a sodium hydroxide (NaOH) solution, leading to a substantial alkaline pH of approximately 12 and, consequently, an ammonia fraction nearing unity; see the yellow shaded region in Figure \ref{figS4}. The ammonia stripping/sampling compartment of the analyzer was structured as shown in Figure 1a.  The stripping column was located at the center of the rack and with a working volume of 0.5 L.  We note that the height of the reservoir (stripping pot) has been designed to be 20 cm to avoid water droplets reaching the top, where the air sampling port is located. A peristaltic sampling pump  was installed above to ensure a proper filling of the stripping column with liquid sample, whereas a peristaltic pump was introducing periodically 0.8 ml of the NaOH solution. Hereafter, an  air pump was activated introducing air required for the ammonia removal, i.e. degassing NH$_3$ into the circulating air. The tail gas, a mixture of air and ammonia, was released through the top of the stripping column and then absorbed by the mSm unit.

\bibliography{biblio}